\documentclass[5p,twocolumn]{elsarticle}  

\usepackage{lineno,float,graphicx,amssymb,amsmath,times,xspace}

\journal{Astroparticle Physics}

\begin{document}

\begin{frontmatter}



\title{Measurement of the cosmic-ray energy spectrum above $10^{16}$ eV with the LOFAR Radboud Air Shower Array}


\author[a]{S. Thoudam\corref{cor1}}
\ead{s.thoudam@astro.ru.nl}
\author[b]{S. Buitink}
\author[a]{A. Corstanje}
\author[a]{J. E. Enriquez}
\author[a,c,d]{H. Falcke}
\author[a,d]{J. R. H\"orandel}
\author[a,e]{A. Nelles}
\author[a]{J. P. Rachen}
\author[a]{L. Rossetto}
\author[a]{P. Schellart}
\author[f,g]{O. Scholten}
\author[a]{S. ter Veen}
\author[f]{T. N. G. Trinh}
\author[a]{L. van Kessel}
\cortext[cor1]{Corresponding author}
\address[a]{Department of Astrophysics, IMAPP, Radboud University Nijmegen, P.O. Box 9010, 6500 GL Nijmegen, The Netherlands}
\address[b]{Astrophysical Institute, Vrije Universiteit Brussel, Pleinlaan 2, 1050 Brussels, Belgium}
\address[c]{ASTRON, 7990 AA Dwingeloo, The Netherlands}
\address[d]{Nikhef, Science Park Amsterdam, 1098 XG Amsterdam, The Netherlands}
\address[e]{Now at: Department of Physics and Astronomy, University of California Irvine, Irvine, CA 92697-4575, USA}
\address[f]{KVI-CART, University of Groningen, 9747 AA Groningen, The Netherlands}
\address[g]{Interuniversity Institute for High-Energy, Vrije Universiteit Brussel, Pleinlaan 2, 1050 Brussels, Belgium}

\begin{abstract}
The energy reconstruction of extensive air showers measured with the LOFAR
Radboud Air Shower Array (LORA) is presented in detail. LORA is a particle
detector array located in the center of the LOFAR radio telescope in the
Netherlands. The aim of this work is to provide an accurate and independent
energy measurement for the air showers measured through their radio signal with
the LOFAR antennas. The energy reconstruction is performed using a
parameterized relation between the measured shower size and the cosmic-ray
energy obtained from air shower simulations. In order to illustrate the
capabilities of LORA, the all-particle cosmic-ray energy spectrum has been
reconstructed, assuming that cosmic rays are composed only of protons or iron
nuclei in the energy range between $\sim2\times10^{16}$ and
$2\times10^{18}$~eV.  The results are compatible with literature values and a
changing mass composition in the transition region from a galactic to an
extragalactic origin of cosmic rays.
\end{abstract}


\end{frontmatter}


\section{Introduction}
\label{sec:intro}
The quest for the origin of cosmic rays is one of the most fundamental problems
in Astroparticle Physics \citep{nagano-watson, Blumer2009, Hoerandel2008}.
Since the discovery of these highly energetic particles more than a century
ago, numerous measurements of several of their properties have been made, using
sophisticated instruments (see e.g. Ref. \citep{Hoerandel2006} for a review).
However, the exact nature of their sources still remains an open question. The
search is mainly hindered due to the fact that cosmic rays, being charged
particles, are scattered or deflected by the Galactic and inter-galactic
magnetic fields during their propagation to the Earth, making it extremely
difficult to reconstruct the direction of their sources. Nevertheless, observed
cosmic-ray properties like the energy spectrum and composition have been used
to understand and characterize the properties of the sources such as their
Galactic or extragalactic nature, the cosmic-ray production spectrum and the
power injected into cosmic rays (see e.g. Refs. \citep{Thoudam-Hoerandel,
Hoerandel2004, Hillas2005, Berezhko2009,Hoerandel2008b, Blasi2013,
Blasi2014} for recent reviews).

LOFAR, the LOw Frequency ARray, is an astronomical radio telescope
\citep{lofar}. It has been designed to measure the properties of cosmic rays
above $\sim10^{16}$~eV by detecting radio emission from extensive air showers
in the frequency range of $10-240$~MHz \citep{Schellart2013}. One of the main
goals of the LOFAR key science project Cosmic Rays is to provide an accurate
measurement of the mass composition of cosmic rays in the energy range between
$\sim10^{16}$ and $\sim10^{18}$~eV, a region where the transition from Galactic
to extragalactic cosmic rays is expected. This is being carried out by
measuring the depth of the shower maximum ($X_\mathrm{max}$), using a technique
based on the reconstruction of the two-dimensional radio intensity profile on
the ground \citep{Buitink2014, Nelles2014}. Another focus of the LOFAR
cosmic-ray measurements is to understand the nature and production mechanisms
of the radio emission from air showers. This is done by measuring various
properties of the radio signals in great detail such as their polarization
properties, the radio wave front and relativistic time compression effects on
the emission profile \citep{Schellart2014, Corstanje2015, Nelles2015}.

In order to assist the radio measurement of air showers with LOFAR, we have
built a particle detector array LORA (LOFAR Radboud Air Shower Array) in the
center of LOFAR \citep{Thoudam2014}. Its main objectives are to trigger the
read-out of the LOFAR radio antennas to register radio signals from air
showers, and to provide basic air shower parameters such as the position of the
shower axis as well as the energy and the arrival direction of the incoming
cosmic-ray.  These parameters are used to cross-check the reconstruction of air
shower properties, based on the measured radio signals.  Currently, given the
lack of an absolute calibration of the radio signals, the cosmic-ray energy is
estimated through the reconstruction of the particle data.  Therefore, an
accurate energy reconstruction with LORA is essential for a proper
understanding of the air showers measured with LOFAR. 

In this article, we describe in detail the various steps of the energy
reconstruction and present the cosmic-ray energy spectrum above
$\sim10^{16}$~eV as measured with LORA. The article is organized as follows. A
short description of the set-up will be given in Section \ref{sec:lora}
followed by a description of the data analysis technique in Section
\ref{sec:analysis}. The various steps involved in the Monte-Carlo simulation
studies of the array will be described in Section \ref{sec:simulation}, and a
comparison between measurements and simulations for some of the air shower
properties will be given in Section \ref{sec:comparison}. In Sections
\ref{sec:energy-calibration} and \ref{sec:energy-resolution}, the energy
calibration, the uncertainties in the reconstructed energies, and reconstructed
cosmic-ray intensity will be described. The measured cosmic-ray spectrum and a
comparison with the measurements of other experiments will be presented in
Section \ref{sec:cosmic-ray-spectrum}, followed by a short conclusion and
a future outlook.

\begin{figure}
\centering
\includegraphics[width=\columnwidth]{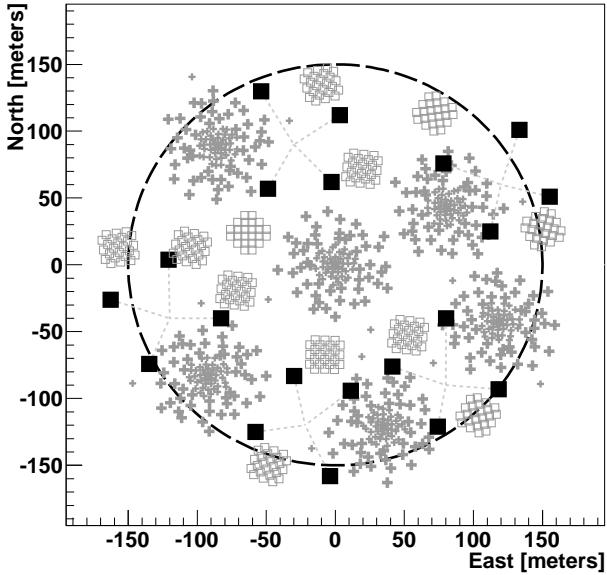}
\caption{Layout of the LORA array in the LOFAR core. The filled black squares
represent the LORA detectors, the crosses the LOFAR low-band antennas and the
empty squares the high-band antennas. The dashed circle in the figure
illustrates the fiducial area of a radius of $150$ m, which is used in the
analysis.}
\label{fig:lora-array}
\end{figure}

\section{LORA experimental set-up and operation}
\label{sec:lora}
LORA (the LOFAR Radboud Air Shower Array) consists of an array of 20 plastic
scintillation detectors of size $\sim0.95~\mathrm{m}\times~0.95$~m each,
distributed over a circular area with a diameter of $\sim320$~m in the center
of LOFAR \citep{Thoudam2014}. The array is subdivided into $5$ units, each
comprising of 4 detectors. The detectors have a spacing between $50-100$~m,
and have been designed to measure cosmic rays with  energies above
$\sim10^{16}$~eV. The array is co-located with six LOFAR stations\footnote{Each
LOFAR station consists of 96 low-band and $48$ high-band antennas, operating
in the frequency range of $10-80$ MHz and $110-240$ MHz respectively.}. The
layout of the array is shown in Figure \ref{fig:lora-array}. The data
acquisition in each unit is controlled locally. A local trigger condition of
3 out of 4 detectors is set for each unit, and an event is accepted for a
read-out of the full array when at least one unit has been triggered. A
high-level trigger for the LOFAR radio antennas is formed when at least 13
out of the 20 detectors have measured a signal above threshold. More
technical details can be found in Ref. \citep{Thoudam2014}.

\section{Data selection and analysis}
\begin{figure*}
\includegraphics[width=\columnwidth]{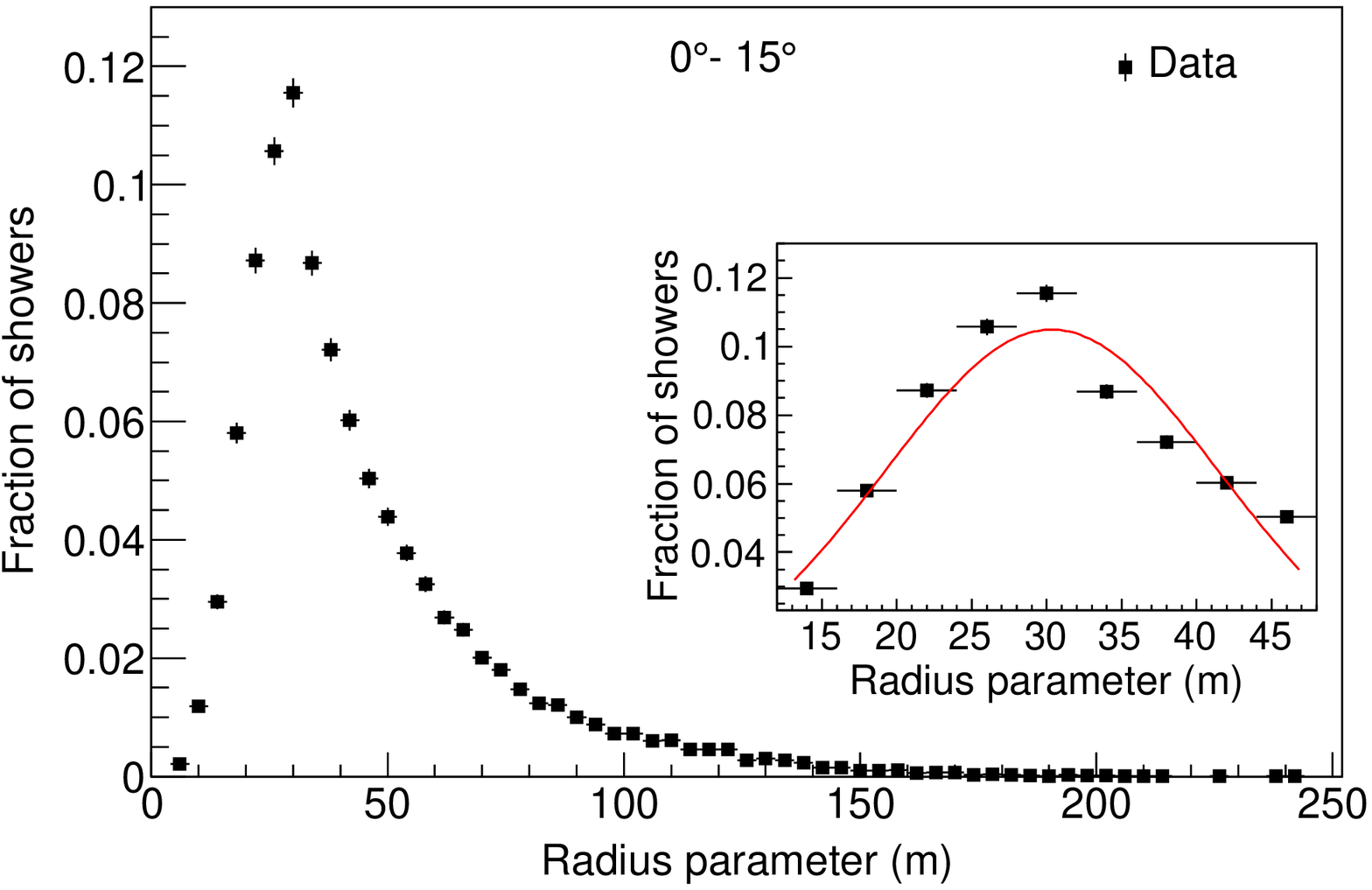}\hspace*{\fill}
\includegraphics[width=\columnwidth]{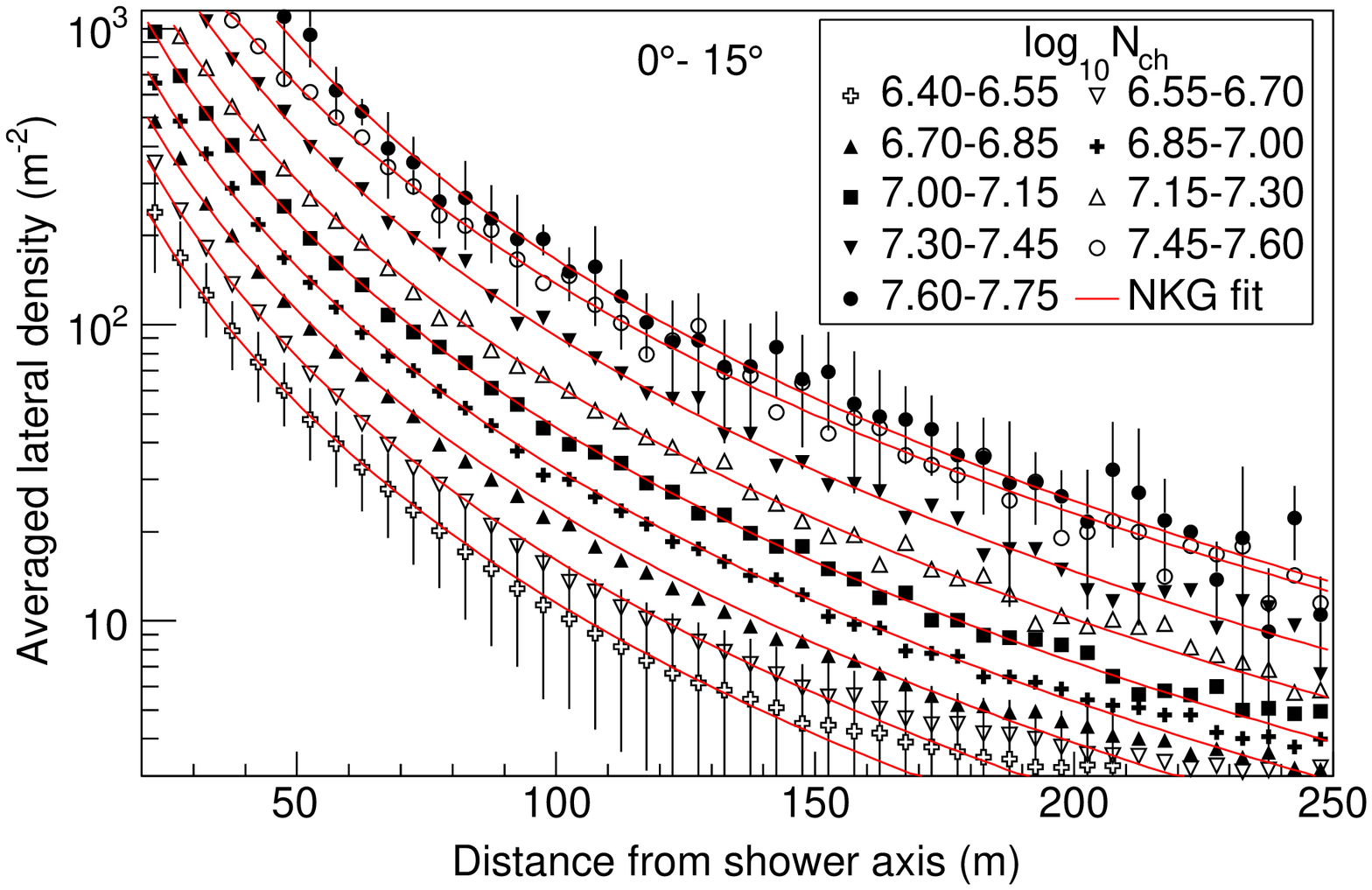}
\caption{\label {average-lateral} Left: Normalized distribution of radius
parameter, $r_\mathrm{M}$, for the measured showers with reconstructed size $\log_{10}N_\mathrm{ch}>6.40$ and zenith angles in the range $0^\circ-15^\circ$. The inset shows a Gaussian fit (represented by
the line) to the distribution around the maximum. Right: Averaged lateral
distribution of measured showers with a reconstructed size in the range of
$6.40<\log_{10}N_\mathrm{ch}<7.75$. Only uncertainties for the uppermost and
the lower-most distribution are shown. The lines represent fits of an NKG
function, keeping the shower age parameter fixed at $s=1.7$.}
\end{figure*}

\label{sec:analysis}
Data collected with the LORA array since its first science operation in June 2011 until October 2014 are used. Only data collected in periods with all 20 detectors in operation will be considered. This amounts to a total of $706.9$ days of data. For the analysis, only showers that trigger a minimum of $5$ detectors will be considered, which corresponds to a total of $1,861,045$ air showers.

For every measured shower, the signal arrival time and the energy deposit in
each detector are recorded. The relative signal arrival times between the
detectors are used to reconstruct the arrival direction of the primary cosmic
ray. The energy deposits are used to reconstruct the position of the shower
axis and the shower size (the effective number of charged particles at the
ground). The latter is determined in terms of the number of vertical equivalent
charged particles, which may also include converted photons in addition to the
dominant charged particles - electrons and muons. The shower axis position and
the shower size are determined simultaneously by fitting a lateral density
distribution function to the measured two-dimensional distribution of particle
densities, projected into the shower plane. The particle density in each
detector is obtained by first dividing the track-length-corrected\footnote{The
measured energy deposit in each detector is corrected for the increase in the
path length of the incident particles through the detector by multiplying by a
$\cos\theta$ factor where $\theta$ is the zenith angle of the reconstructed
arrival direction of the primary cosmic ray.} energy deposition by the energy
deposition of a single particle obtained from calibration, and then by further
dividing by the projected area of the detector in the shower plane. The lateral
density distribution of an air shower is generally described by the
Nishimura-Kamata-Greisen (NKG) function which is given by \citep{Kamata1958,
Greisen1960}
\begin{equation}
\label{eq:lateral-density}
\centering
\rho(r)=N_\mathrm{ch} C(s) \left(\frac{r}{r_\mathrm{M}}\right)^{s-2}\left(1+\frac{r}{r_\mathrm{M}}\right)^{s-4.5},
\end{equation}
where $\rho(r)$ represents the particle density in the shower plane at a radial distance $r$ from the shower axis, $N_\mathrm{ch}$ is the shower size, $s$ is shower age or lateral shape parameter and $r_\mathrm{M}$ is the radius parameter which is basically a measure of the lateral spread of the shower. The function $C(s)$ is given by
\begin{equation}
\label{eq-Cs}
\centering
C(s)=\frac{\Gamma(4.5-s)}{2\pi r_\mathrm{M}^2 \Gamma(s)\Gamma(4.5-2s)} .
\end{equation}
In the case of LORA, the value of $r_\mathrm{M}$ is determined from the fit along with $N_\mathrm{ch}$ and the position of the shower axis. The parameter $s$ is kept constant at a value of $1.7$ throughout the fitting process. Simultaneous fitting of both $r_\mathrm{M}$ and $s$ results in fits of poorer quality. Simulation studies have shown that keeping $s$ constant gives better results than keeping $r_\mathrm{M}$ constant \citep{Antoni2001}. The fitting procedure is repeated three times with the output of each fit taken as starting values for the next iteration. Details about the minimization procedure and the choice of starting values, as well as the reconstruction of the arrival direction of the primary particle are described in Ref. \citep{Thoudam2014}.

All showers that trigger at least $5$ detectors with a minimum of
$1$~particle~m$^{-2}$ are allowed to pass through the reconstruction algorithm,
and their shower parameters are calculated. Furthermore, only showers whose
reconstructed position of the shower axis falls within $150$~m from the center
of the array are selected. The normalized distribution of $r_\mathrm{M}$ values
for the selected showers with reconstructed sizes $\log_{10}N_\mathrm{ch}>6.40$ and reconstructed zenith angles in the range of
$0^\circ-15^\circ$ are shown in Figure \ref{average-lateral} (left panel). The
inset shows a closer view of the distribution around the maximum value between
$12$ and $48$~m, and a Gaussian fit to the distribution. The fit gives a peak
value of $r_\mathrm{M}=30.33\pm0.13$~m. Figure \ref{average-lateral} (right
panel) shows the averaged lateral distributions of the measured showers for
different reconstructed size bins in the range of
$6.40<\log_{10}N_\mathrm{ch}<7.75$ for zenith angle between $0^\circ$ and
$15^\circ$. The distributions include events that passed through the same
selection cuts applied in the left panel of Figure \ref{average-lateral} and
have $r_\mathrm{M}$ values in the range of $10-200$~m. The averaged
distributions are obtained by stacking together the lateral distributions of
all individual showers contained in each size bin. The lines in the plot
represent the fits to the data using (\ref{eq:lateral-density}). To
avoid clumsiness of the plots, uncertainties are shown only for the size bins
of $\log_{10}N_\mathrm{ch}=6.40-6.55$ and $\log_{10}N_\mathrm{ch}=7.60-7.75$.
However, all the respective uncertainties are taken into account in the fitting
procedure. The values of $r_\mathrm{M}$ obtained from the fits are in the range
of $23-31$~m.

\begin{figure}
\centering
\includegraphics[width=\columnwidth]{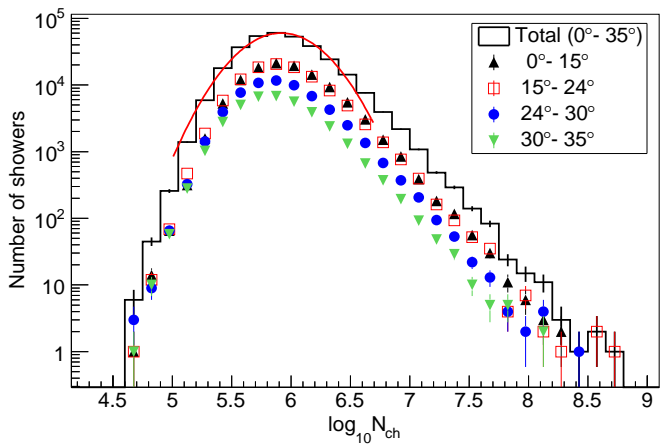}
\caption{\label {size-distribution} Measured size distribution of the air showers that have passed the quality cuts given in Table \ref{quality-cut}. The line represents a Gaussian fit to the total distribution around the peak, giving a shower size threshold of $\log_{10}N_\mathrm{ch}=5.92$ for the array.}
\end{figure}

The shower size gives a good measure of the energy of the primary cosmic-ray
particle, initiating the air shower. Therefore, the shower size distribution
should reflect the energy distribution of the cosmic rays at size values where
the primary energy is above the detector threshold. Figure
\ref{size-distribution} shows the distribution of reconstructed shower sizes
for all the measured showers that passed through the various trigger and
quality cuts applied in the analysis. This corresponds to a total of $322,664$
air showers. The distribution shows a steep rise as $N_\mathrm{ch}$ increases
which is due to the sharp increase in the detector acceptance (see section
\ref{sec:simulation}) as function of the primary energy. After reaching a
maximum, the distribution falls off steeply which is due to the power-law
behavior of the cosmic-ray spectrum. The peak of the total distribution gives
the shower size threshold of the detector array. Fitting a Gaussian function around the
peak gives a value of $\log_{10}N_\mathrm{ch}=5.92$. Also shown in Figure
\ref{size-distribution} are the reconstructed size distributions for
four zenith angle bins: $0^\circ-15^\circ$, $15^\circ-24^\circ$,
$24^\circ-30^\circ$, and $30^\circ-35^\circ$. The parameterization of the
cosmic-ray energy will be determined separately for each zenith angle bin (see
Section \ref{sec:energy-calibration}). All cuts applied in this analysis are
summarized in Table \ref{quality-cut}.

\begin{table*}[t]
\centering
\caption{Selection cuts applied in the analysis of both the measurements and
the air shower simulations.}
\vspace{\baselineskip}
\label{quality-cut}
\begin{tabular}{ll}
\hline
 \textbf{Trigger condition:} & \\
 Single unit trigger: & 3/4 detectors \\
 Analysis: & 5 detectors with $\ge1$~particle~m$^{-2}$\\
 Number of leftover showers: &$1,861,045$\\
 \hline
\textbf{Quality cuts:} & \\
 Zenith angle: & $\theta<35^\circ$ \\
 Position of the shower axis: & $<150$~m from array center\\
 Radius parameter: & $10~\mbox{m} < r_\mathrm{M} < 200~\mbox{m}$ \\
 Number of leftover showers: &$322,664$\\
 \hline
\end{tabular} 
\end{table*}

\section{Simulations}
\label{sec:simulation}
Detailed simulation studies have been carried out in order to understand the
performance of the array and to determine various characteristics of the array,
such as the trigger and reconstruction efficiencies, the reconstruction
accuracies of shower parameters, the relation between reconstructed size and
primary energy, and the accuracy in the energy reconstruction. In this section,
the various steps involved in the simulations will be described.

\subsection{Air shower simulations}
\label{subsec:air-shower}
Air showers are simulated using the CORSIKA simulation package (version
$7.4387$) \citep{Heck1998}. The interactions of hadronic particles in the
Earth's atmosphere are treated using QGSJET-II-04 \citep{Ostapchenko2011} at
high energies and FLUKA \citep{Fasso2005} for energies below $200$~GeV. The
electromagnetic interactions are treated with EGS4 \cite{Nelson1985}. The
observation level of the LORA array is set to $7.6$~m above sea level. Air
showers are simulated for protons and iron nuclei in the energy range of
$10^{15}-10^{19}$~eV, assuming a differential energy spectrum with an index
$-2$.  The showers are weighted to generate a distribution with a spectral
index $-3$. Zenith angles are considered in the range  $0^\circ-45^\circ$. In
order to reduce the excessive computing times involved in generating the
showers, `thinning' is applied at a level of $10^{-6}$ with optimized weight
limitation \citep{Kobal2001}.

\subsection{Detector simulation}
\label{detector-simulation}
The generated air shower particles are fed into a detector simulation code,
based on the GEANT4 package \citep{Agostinelli2003}, which allows to calculate
the total energy deposition in each detector. All properties of the detector,
such as the type and the density of the scintillator material, the detector
geometry as well as the effect of the aluminum plates covering the scintillator
plates are included in the simulation. In order to avoid air showers not
creating a trigger in the detectors due to the large detector spacing of the
LORA array, an additional step is applied to each simulated shower before
feeding the particles into GEANT4. Concentric rings with a radial bin size of
2~m centered around the shower axis are constructed, and the total number of
particles contained in each projected ring on the ground is calculated. All
particles in a ring are then distributed uniformly in a small square region of
area $A_\mathrm{s}=(1.5\times1.5)$~m$^2$ with a LORA detector in its
center.  Depending on the arrival direction of the particles, those that hit
the detector are allowed to pass through GEANT4 and the total energy deposition
$E_\mathrm{dep}$ in the detector is obtained. In the final step, the actual
amount of energy that would have been deposited in the detector is obtained by
applying a correction
$E'_\mathrm{dep}=E_\mathrm{dep}A_\mathrm{s}\cos\theta/A_\mathrm{R}$, where
$\theta$ is the zenith angle of the shower and $A_\mathrm{R}/\cos\theta$ is the
projected area of the ring on the ground. The somewhat larger area of
$A_\mathrm{s}$ than the actual detector area is used to accommodate particles
hitting the detector at larger zenith angles. For each simulated shower, the
radial distribution of the energy deposition in the detector, averaged over the
azimuthal direction in the shower plane, is constructed as a function of the
distance to the shower axis. This method also automatically allows to correct
for the effect of the shower thinning applied in CORSIKA as the calculation
takes into account all the particles arriving at the ground.

\begin{figure*}
\includegraphics[width=\columnwidth]{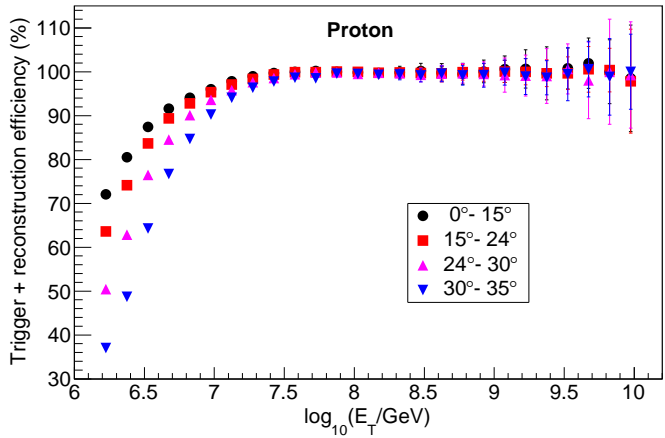}
\hspace*{\fill}
\includegraphics[width=\columnwidth]{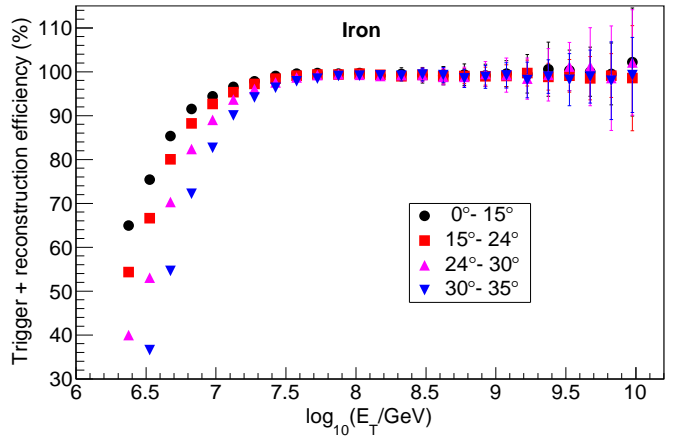}
\caption{\label {total-efficiency} Combined trigger and reconstruction
efficiencies obtained from the simulation of  showers induced by protons (left)
and iron nuclei (right) as a function of the true energy. Different symbols
represent different zenith angle bins. See Section \ref{sec:efficiency} for
details.}
\end{figure*}

Simulations have also been performed to calculate the energy deposition of
singly charged particles in the detector. For that, muons of an energy of 4~GeV
are considered. Energy depositions for vertical incident muons
and for muons following a realistic (observed) arrival direction distribution
are obtained. The energy deposition distribution for vertical muons gives a
most probable value of $E_\mathrm{VEM}=5.3$~MeV, while the all-sky distribution
gives $6.67$~MeV.  The latter is obtained by also taking into account a noise
level of $\sigma=1$~MeV, which includes a contribution from statistical noise,
generated by the low number of scintillation photons producing a signal and the
electronic noise. The energy deposition for the all-sky distribution is used to
calibrate the distribution of the total energy deposition by single particles
measured with the experiment. Details about the calibration are described in
Ref. \citep{Thoudam2014}.

\begin{figure}
\centering
\includegraphics[width=\columnwidth]{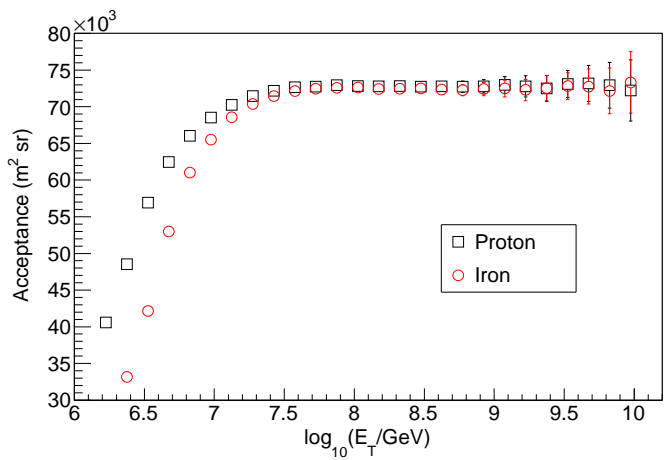}
\caption{\label {total-effective-area} Total acceptance for showers induced by
protons (squares) and iron nuclei (circles) obtained from simulations as
function of the true energy. The acceptance is calculated for solid angles
subtended within $0^\circ-35^\circ$. See Section \ref{sec:efficiency} for
details.}
\end{figure}

\subsection{Reconstruction of shower parameters}
Every simulated shower is assigned a random position on the ground. The
position of the shower axis, and also the detector coordinates, are then
projected in the shower plane. Based on the distance of the detector from the
position of the shower axis in the shower plane, the amount of energy deposited
in the detector is calculated from the radial distribution of energy deposition
given by the simulation. To make the simulation study consistent with the
analysis of the measured data, the number of particles hitting the detectors is
obtained in units of VEM (vertical equivalent muons). This is done by first
dividing the track-length-corrected energy deposition
$E'_\mathrm{dep}\cos\theta$ by $E_\mathrm{VEM}$ to obtain the mean number of
VEM particles $\bar{n}$, hitting the detector. To obtain a realistic value, the
detector is assigned a number, drawn randomly from a Poisson distribution with
mean $\bar{n}$. This last step is necessary to correct for the azimuthal
averaging of the energy depositions around the shower axis, applied in the
simulation. The final value $n_\mathrm{f}$ for the number of VEM particles is
obtained by adding a random noise, drawn from a Gaussian distribution with a
standard deviation $\sigma/E_\mathrm{VEM}$. The particle density in each
detector is obtained by dividing $n_\mathrm{f}$ by the projected area of the
detector $A_\mathrm{d}\cos\theta$, where $A_\mathrm{d}$ is the actual
geometrical area of the detector. After obtaining the particle densities in the
detectors, the reconstruction of air shower parameters is performed similar to the reconstruction of the measured air shower data.

\subsection{Trigger and reconstruction efficiencies}
\label{sec:efficiency}
In order to improve the statistics, each simulated shower is processed $100$
times with the position of the shower axis selected randomly within a circle
with a radius of 160~m from the center of the array. The fiducial cut of 150~m
applied in the data analysis is also applied in the calculation of the detector
efficiency. A larger radius of 160~m with respect to the fiducial cut is
necessary to take into account the spillover of reconstructed showers across
the fiducial boundary due to the limited reconstruction accuracy in the
position of the shower axis which reaches $\sim10$~m at a distance of 150~m
from the array center. Only showers with zenith angles within
$0^\circ-35^\circ$ are considered, and are divided into four different zenith
angle bins as in the data analysis. For each energy and zenith angle bin, the
trigger efficiency, $\epsilon_\mathrm{t}$, is determined by taking the ratio of
the number of showers that pass through the trigger condition listed in Table
\ref{quality-cut} to the total number of showers generated with true shower
axis position within the fiducial area. The reconstruction efficiency, 
$\epsilon_\mathrm{r}$, is calculated as the ratio of the number of showers that
pass through both the trigger and the quality cuts to the total number of
triggered showers. Then, the total efficiency is obtained as,
$\epsilon_\mathrm{tot}=\epsilon_\mathrm{t}~\epsilon_\mathrm{r}$. Figure
\ref{total-efficiency} shows the total efficiency for protons (left panel) and
iron nuclei (right panel) as a function of the true energy $E_\mathrm{T}$ for
the four zenith angle bins: $0^\circ-15^\circ$, $15^\circ-24^\circ$,
$24^\circ-30^\circ$, and $30^\circ-35^\circ$.  The full efficiency of $100\%$ is
reached at $\log_{10}(E_\mathrm{T}/\mathrm{GeV})\approx7.6$ for protons  and
at $\approx7.7$ for iron nuclei.

\begin{figure*}
\includegraphics[width=\columnwidth]{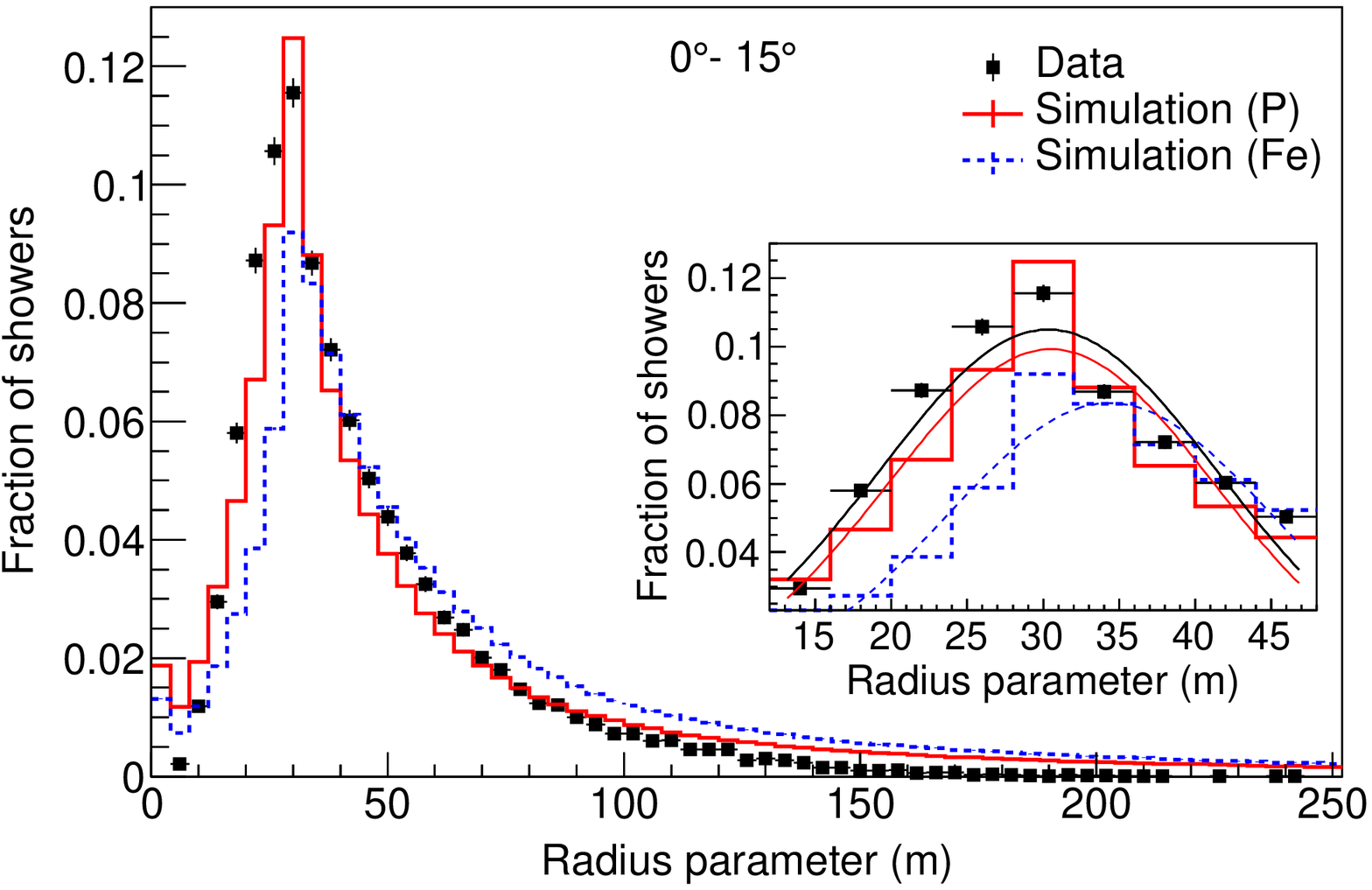}
\hspace*{\fill}
\includegraphics[width=\columnwidth]{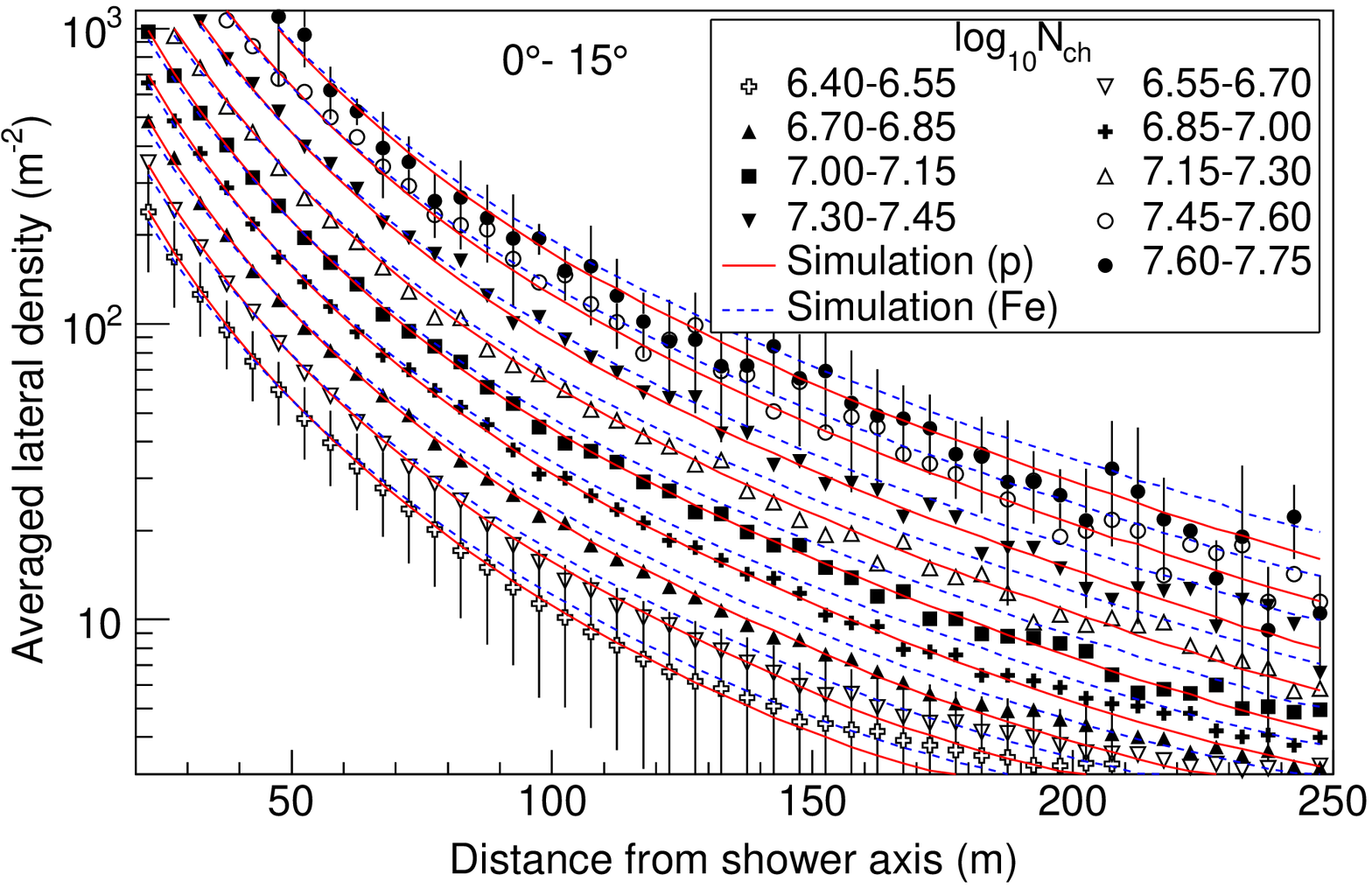}
\caption{\label {average-lateral-data-sim} Left: Comparison of normalized
distributions of radius parameter obtained from the measurements (points) and
simulations (thick-solid line: protons and thin-dashed line: iron nuclei) for showers with reconstructed sizes $\log_{10}N_\mathrm{ch}>6.40$ and reconstructed zenith angles between $0^\circ$ and $15^\circ$. The inset shows Gaussian fits (thin lines) to the distributions around the maximum. Right: Comparison of the
averaged lateral distributions between measurements and simulations. The
measurements (symbols) are the same as shown in Figure \ref{average-lateral}
right panel and the lines (solid: protons and dashed: iron nuclei) are the
simulation results for the same shower size bin.}
\end{figure*}

Figure \ref{total-effective-area} shows the total acceptance of the array for
primary protons and iron nuclei as a function of the true energy. The detector
acceptance $A_\mathrm{acc}$ is defined as the total effective area of the array
multiplied by the effective viewing angle, and it is calculated as,
\begin{equation}
\label{eq:acceptance}
A_\mathrm{acc}(E_\mathrm{T})=\int^{\Omega_c}_0 A_\mathrm{proj}(\theta)\epsilon_\mathrm{tot}(E_\mathrm{T},\theta,\phi) d\Omega,
\end{equation}
where $d\Omega=\sin\theta~d\theta~d\phi$ is the solid angle subtended by an element of opening angle between $\theta$ and $\theta+d\theta$ and an azimuthal width of $d\phi$. $\Omega_c$ is the maximum solid angle corresponding to the zenith angle cut of $\theta_c=35^\circ$, $A_\mathrm{proj}=\pi~R_\mathrm{c}^2 ~\cos\theta$ is the projected geometrical area of the array at an inclination $\theta$ with $R_\mathrm{c}=150$~m representing the fiducial radial cut applied in the analysis, and the total efficiency, $\epsilon_\mathrm{tot}$, is given as a function of $(E_\mathrm{T}, \theta, \phi)$. Assuming azimuthal symmetry of $\epsilon_\mathrm{t}$, the integral in Equation \ref{eq:acceptance} is discretized in zenith angle bins and can be rewritten as,
\begin{equation}
A_\mathrm{acc}(E_\mathrm{T})=\frac{\pi^2 R^2_\mathrm{c}}{2}\sum\limits_{k=1}^{n_\theta}
\epsilon_\mathrm{t}(E_\mathrm{T},\theta_\mathrm{k})\\ 
\left(\cos2\theta_\mathrm{k}-\cos2\theta_\mathrm{k+1}\right),
\end{equation}
where $k$ denotes the zenith angle bins, $n_\theta=4$ is the number of zenith
angle bins considered, and $\theta_\mathrm{k}$ and $\theta_\mathrm{k+1}$
represent the low-bin and high-bin edges of each zenith angle bin respectively.

\section{Comparison between simulations and measurements}
\label{sec:comparison}

In Figure \ref{average-lateral-data-sim} (left panel), the normalized distribution of radius parameters for the simulated showers with reconstructed sizes $\log_{10}N_\mathrm{ch}>6.40$ is compared with the measurements for the zenith angle range of $0^\circ-15^\circ$. The points in the figure represent the measurements and they are the same as shown in Figure \ref{average-lateral} (left panel). The
distribution for iron nuclei (thick-dashed line) shows a systematic shift
towards larger $r_\mathrm{M}$ with respect to the proton showers (thick-solid
line) which is expected due to the difference in the shower development between
proton and iron primaries. Showers induced by iron nuclei are generated higher up in the
atmosphere, resulting in a larger spread (which implies larger $r_\mathrm{M}$
values) on the ground, relative to the proton showers. Although both the simulated
distributions follow a similar shape as the measured distribution, they are not
in full agreement with the data. But, overall, the proton distribution seems to be relatively closer to the data. The inset shows a closer view for the region around the maximum between 12 and 48~m. The lines represent fits to the distributions using a Gaussian function. The Gaussian peaks for the simulated distributions obtained from the fits are
$(30.51\pm0.01)$~m for the proton distribution and $(34.38\pm0.02)$~m for the
iron distribution. The value for the proton distribution is found to be quite close to the peak value of $(30.33\pm0.13)$~m obtained for the data. 

Figure \ref{average-lateral-data-sim} (right panel) shows a comparison of the averaged lateral distribution between simulations and measurements for a reconstructed shower size in the range of $6.40<\log_{10}N_\mathrm{ch}<7.75$. The measurements (points) are the same as already shown in Figure \ref{average-lateral} (right panel). The iron distributions (dashed lines) are  found to be slightly flatter than the proton distributions (solid lines), which is expected due to the larger $r_\mathrm{M}$ values for iron showers as explained above. Although both the simulated proton and iron distributions are consistent with the data within the experimental uncertainties, the proton distributions seem to agree better as the iron distributions tend to show some systematic deviation from the measurements above a distance of $\sim70$~m from the shower axis. A $\chi^2$ test of the comparison between the simulations and the measurements gives reduced $\chi^2$ values within the range of $\sim 1.15-1.26$ for protons and $\sim 1.33-1.86$ for the case of iron nuclei. The better agreement of the measurements with the proton distributions is expected because the air shower particles measured by LORA are mostly dominated by electrons rather than muons, which makes the measurements biased towards protons.

\begin{figure*}
\includegraphics[width=\columnwidth]{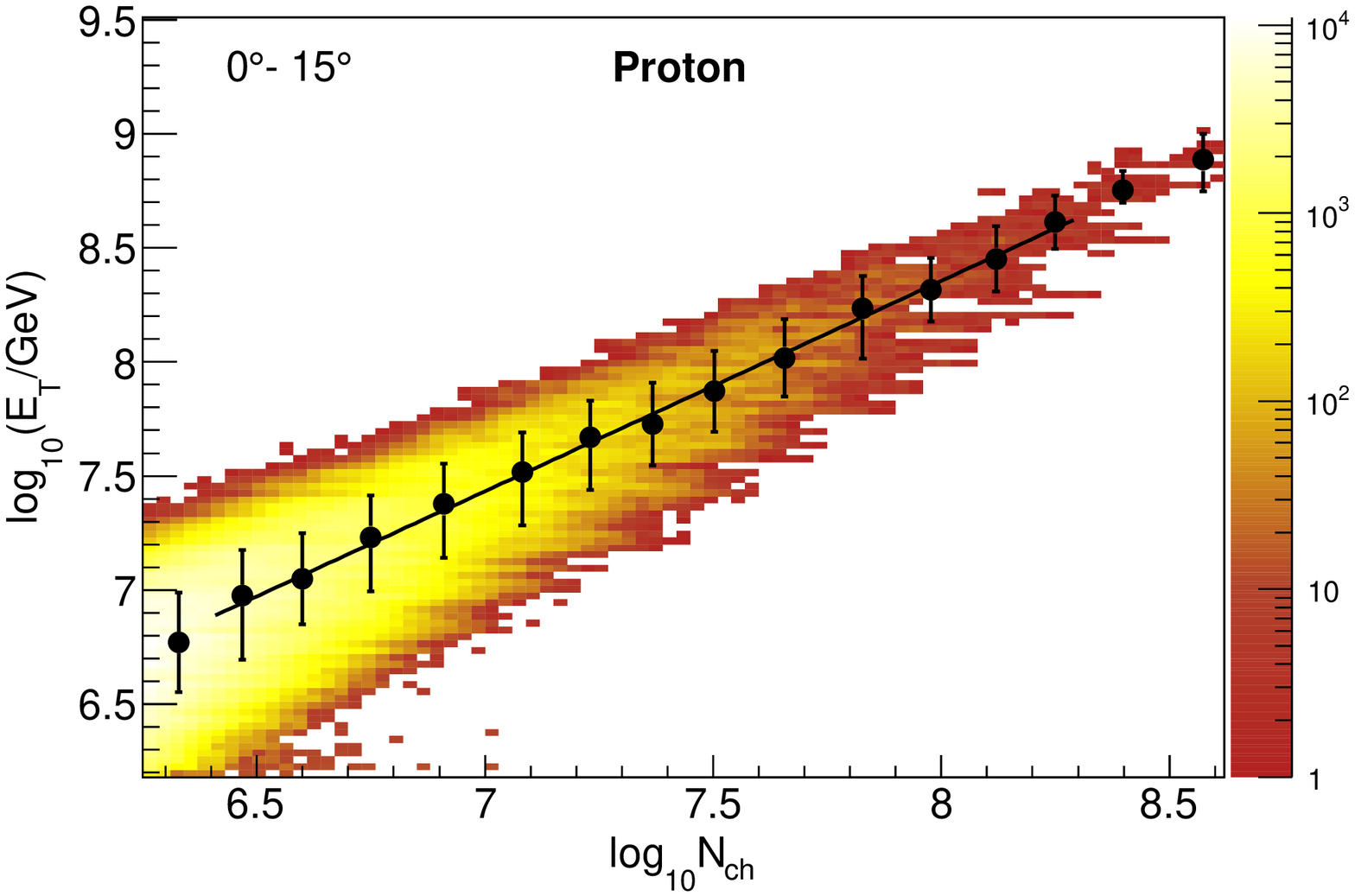}
\hspace*{\fill}
\includegraphics[width=\columnwidth]{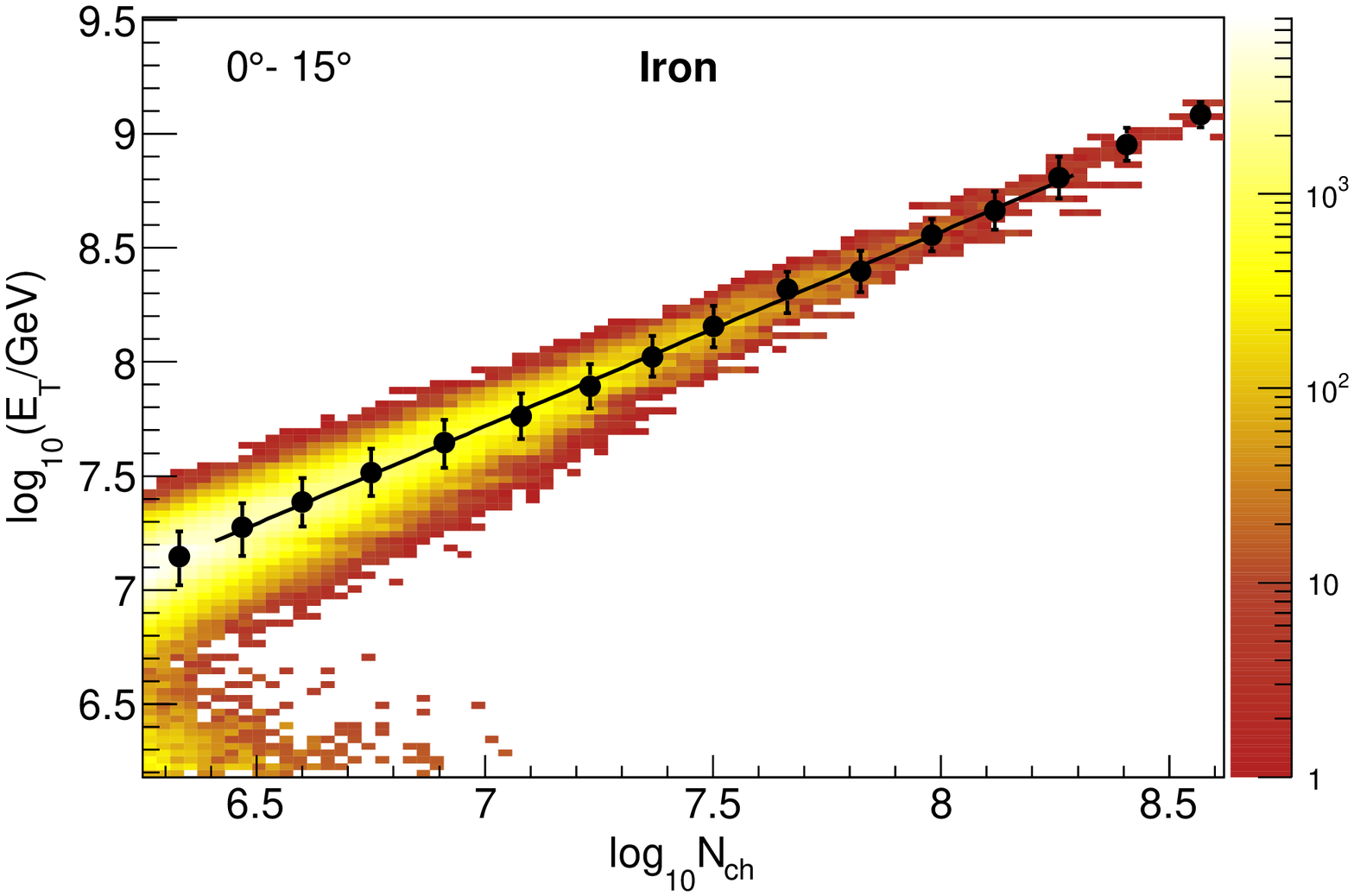}
\caption{\label{size-energy-plot} 
Two-dimensional histogram for the reconstructed shower size $N_\mathrm{ch}$ and
true energy $E_\mathrm{T}$, obtained from simulations for showers induced by
protons (left) and iron nuclei (right) for a zenith angle bin of
$0^\circ-15^\circ$. The distributions are weighted to an energy spectrum of index -3. Each point represents the peak in the true energy distribution for a $\log_{10}N_\mathrm{ch}$ bin width of 0.15 (see Section
\ref{sec:energy-calibration} for details). The lines represent fits to the
points using Equation \ref{eq:size-energy} in the range of
$6.4<\mathrm{log}_{10}N_\mathrm{ch}<8.3$ for protons and
$6.3<\mathrm{log}_{10}N_\mathrm{ch}<8.3$ for iron nuclei. The fit parameters are listed in Table 2.}
\end{figure*}

\begin{figure*}
\includegraphics[width=\columnwidth]{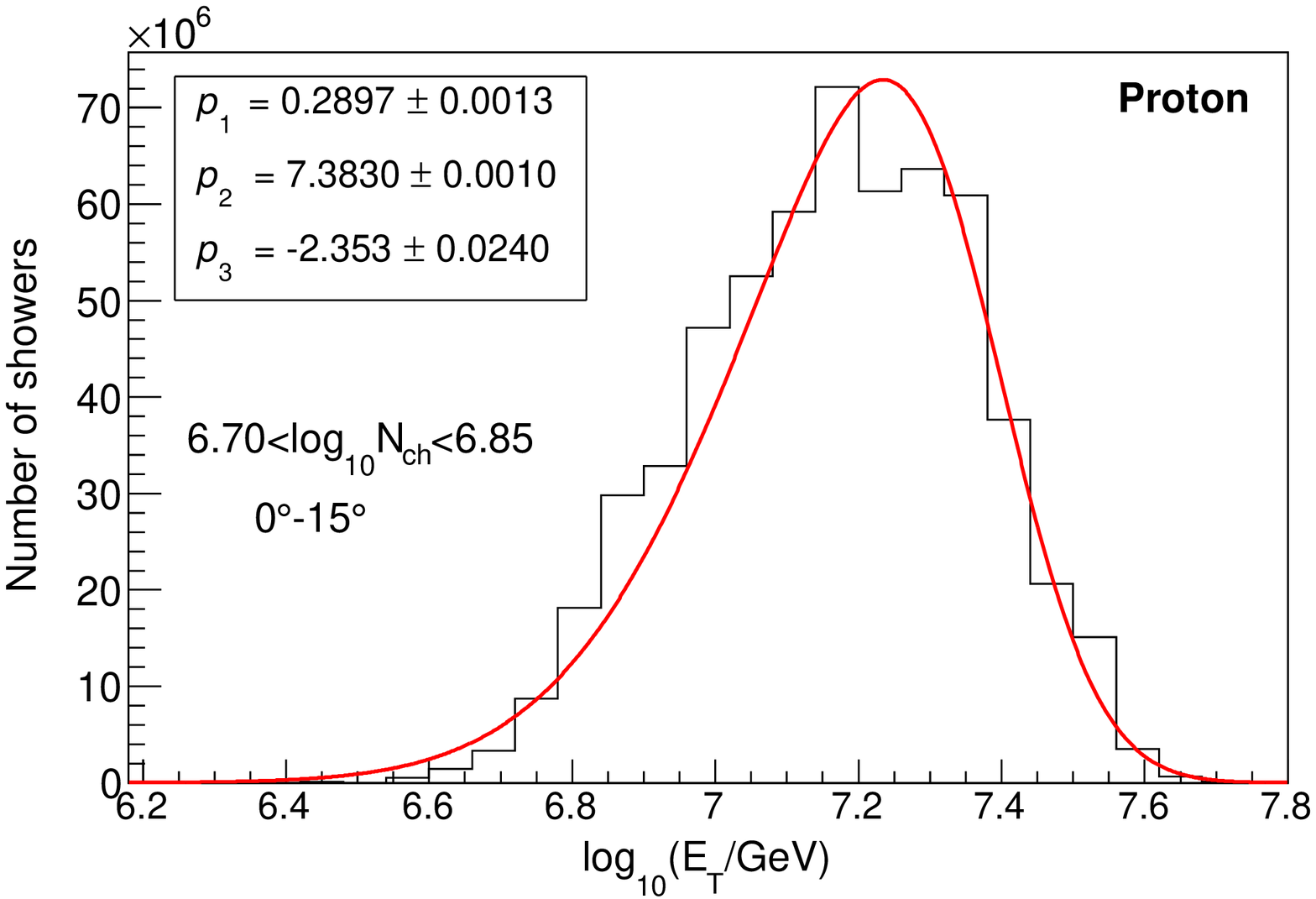}
\hspace*{\fill}
\includegraphics[width=\columnwidth]{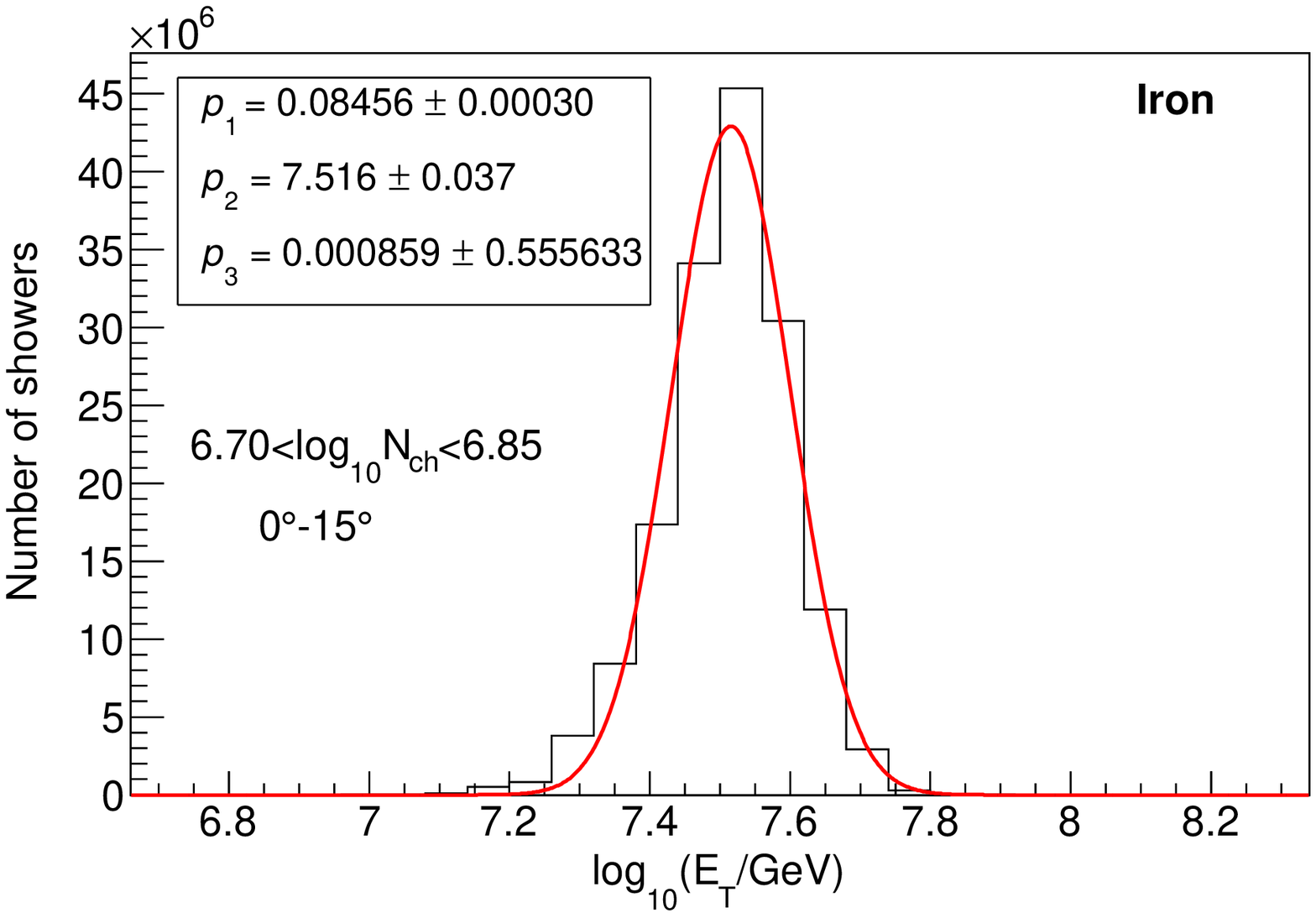}
\caption{\label {energy-dis} True energy distribution for the reconstructed
size bin of $6.70<\mathrm{log}_{10}N_\mathrm{ch}<6.85$ for protons (left panel) and
iron nuclei (right panel) for the zenith range $0^\circ-15^\circ$. The distributions are weighted to an energy spectrum of index -3. The lines represent fits using a skewed Gaussian function given by Equation \ref{eq:skewed-gaussian} which involves four parameters (see Section \ref{sec:energy-calibration} for details). Main fit parameters $(p_1, p_2, p_3)$ are shown.}
\end{figure*}

\begin{table*}
\centering
\caption{Fit parameters for showers induced by protons and iron nuclei obtained
by fitting Equation \ref{eq:size-energy} to the size-energy profile plots for
different zenith angle bins between $0^\circ$ and $35^\circ$. A slope
$\gamma_\mathrm{s}=-3$ of the differential cosmic-ray energy spectrum has
been adopted in the simulations.}
\vspace{\baselineskip}
\label{fit-parameters}
\begin{tabular}{cccccc}
\hline
\multicolumn{1}{c}{Zenith angle} & \multicolumn{2}{c}{Protons}&& \multicolumn{2}{c}{Iron nuclei} \\
\cline{2-3} \cline{5-6}
$\theta$&a&b&&a&b\\
\hline
$0^\circ-15^\circ$ & $0.980\pm0.683$ & $0.922\pm0.089$&& $1.747\pm0.361$ & $0.853\pm0.048$ \\
$15^\circ-24^\circ$ & $1.234\pm0.766$ & $0.898\pm0.099$&& $1.801\pm0.370$ & $0.858\pm0.049$ \\
$24^\circ-30^\circ$ & $1.315\pm0.815$ & $0.901\pm0.101$&& $1.726\pm0.319$ & $0.885\pm0.041$ \\
$30^\circ-35^\circ$ & $1.667\pm0.692$ & $0.873\pm0.091$&& $1.982\pm0.366$ & $0.866\pm0.048$ \\
\hline
\end{tabular}
\end{table*}

\section{Energy calibration}
\label{sec:energy-calibration}
The measured shower size can be converted into the energy of the primary
particle using a conversion relation obtained from simulations. Simulated
showers are stored in a two dimensional log-log histogram in reconstructed
size and true energy. Such a histogram is shown in Figure
\ref{size-energy-plot} for showers induced by protons (left panel) and iron
nuclei (right panel) for the zenith angle range of $0^\circ-15^\circ$.  The
color profile represents the weight of the distribution. The distribution is
broader for proton showers, which is mainly due to the large intrinsic
fluctuations of proton showers. From simulations, the fluctuations in the true
shower size for proton showers within the $0^\circ-15^\circ$ zenith angle bin
are found to be $\sim(30-45)\%$ while the uncertainty due to the reconstruction
is in the range of $\sim(12-19)\%$ for the energy region of our interest. For
iron induced showers, the intrinsic size fluctuation is only $\sim(17-22)\%$,
while the reconstruction accuracy remains almost the same as that of the proton
induced showers. Another major difference is that for the same reconstructed
shower size, iron showers have higher energies than the protons. This is
related to the shallower penetration depth of iron induced showers in the
atmosphere, which leads to an increased attenuation of electrons before they can
reach the ground.

The distributions in Figure \ref{size-energy-plot} are binned in
$N_\mathrm{ch}$, taking a logarithmic bin size of $0.15$, and profile plots of
the true energy as function of $N_\mathrm{ch}$ are generated. The profile plots
are represented by the solid points in Figure \ref{size-energy-plot}. Each
point in the plots represents the peak of the energy distribution for each size
bin, and the uncertainty on each point corresponds to the spread of the energy distribution which
is described in detail in the following. Figure \ref{energy-dis} shows the
energy distribution for the bin slice  of $\log_{10}N_\mathrm{ch}=6.70-6.85$
for both, showers induced by protons (left panel) and iron nuclei (right
panel). The distribution for protons is not symmetric about its mean and is
found to be more extended to lower energies. This can be understood as more
contamination from low-energy showers in a given size bin than from higher
energies which is caused by the larger intrinsic fluctuations of low-energy
showers. The level of contamination depends on the assumed slope of the primary
cosmic-ray spectrum in the simulation. The peaks of the distributions are
obtained by fitting with a skewed Gaussian function. The skewed Gaussian
distribution function used in the present analysis is given by, 
\begin{equation}
\label{eq:skewed-gaussian}
f(x)=\frac{p_0}{p_1}\exp\left(-\frac{\left(x-p_2\right)^2}{2p_1^2}\right)\left[1+\mathrm{erf}\left(\frac{p_3(x-p_2)}{\sqrt{2p_1^2}}\right)\right],
\end{equation} 
where $x=\log_{10}\left(E_\mathrm{T}/\mathrm{GeV}\right)$, $p_0$ is the normalisation constant, $p_1$ and $p_2$ represent measures of the spread and the position of the distribution respectively, and $p_3$ is the skewness parameter of the distribution function.

\begin{figure*}
\includegraphics[width=\columnwidth]{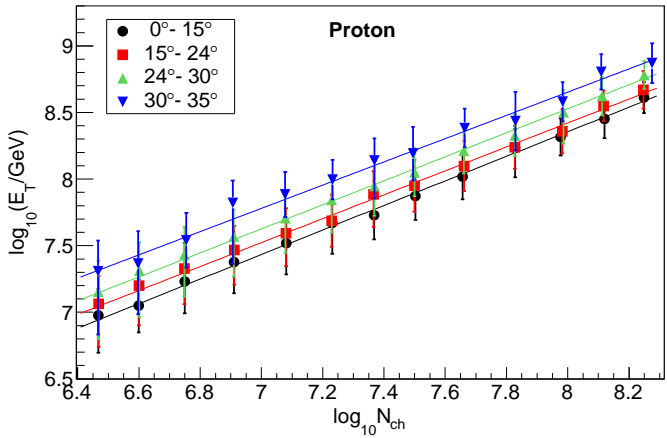}
\hspace*{\fill}
\includegraphics[width=\columnwidth]{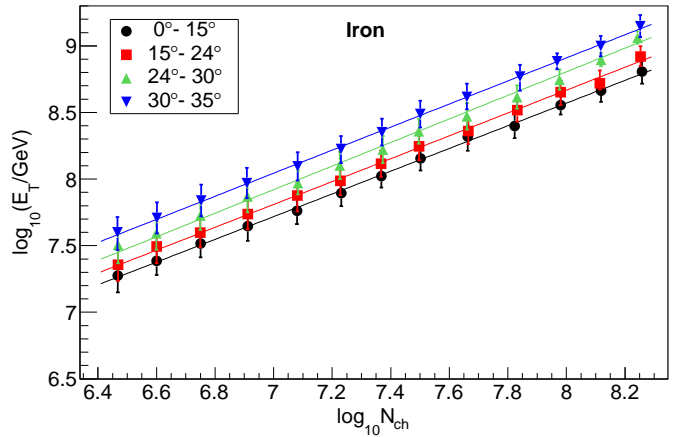}
\caption{\label {size-energy-all-angles} Reconstructed size ($N_\mathrm{ch}$) and true energy ($E_\mathrm{T}$) relation for proton (left) and iron (right) showers for all zenith angle ranges up to $35^\circ$.}
\end{figure*}

The lines in Figure \ref{energy-dis} represent the fitted functions. The important fit parameters $(p_1, p_2, p_3)$ are also shown. For the distribution of iron-induced showers, it can be noticed that the value of $p_3$ is close to zero, indicating
that the distribution closely resembles a normal Gaussian distribution. The
uncertainties in the profile plots shown in Figure \ref{size-energy-plot} are obtained by taking the difference between the
energies corresponding to the full width at half maximum (FWHM) and the peak energy of the energy distribution for each size bin. The uncertainties obtained are asymmetric for the proton distribution while for iron induced showers, they are
almost symmetric. The $N_\mathrm{ch}$ values of the profile plots shown in
Figure \ref{size-energy-plot} are calculated as the weighted mean of the size
distribution within each size bin. These size values are found to be slightly
smaller than the bin centers.

To obtain the size-energy relation, each profile plot is fitted using the following function,
\begin{equation}
\label{eq:size-energy}
\centering
\log_{10}E_\mathrm{T}=a+b\log_{10}N_\mathrm{ch}
\end{equation}
where $E_\mathrm{T}$ denotes the true energy, and $a$ and $b$ are the fit
parameters. The fit is performed only in the size region where a reliable fit
of the true energy distribution, as shown in Figure \ref{energy-dis}, could be
performed. This corresponds to a size region of $\log_{10}N_\mathrm{ch}=6.4-8.3$ for both the type of particles. The profile plots as well as the fitted functions for all the zenith angle ranges
are shown in Figure \ref{size-energy-all-angles} for protons (left panel) and
iron nuclei (right panel).

From these figures, it can be noticed that for the same shower size, primary
energies at larger zenith angles are larger than at smaller angles. In other
words, it requires a higher energy at larger zenith angles to generate the same
number of particles on the ground as at lower zenith angles. This is due to
higher attenuation of air shower particles at larger zenith angles as the
showers pass through a longer column depth of air in the atmosphere. The values
of the $a$ and $b$ parameters obtained from the fits for the four zenith angle
ranges are listed in Table \ref{fit-parameters}. Using these values, for any
simulated or measured shower for which the reconstructed arrival direction and
the reconstructed size are known, the primary cosmic-ray energy can be
reconstructed using the relation
\begin{equation}
\centering
\label{eq:energy-calibration}
\log_{10}E_\mathrm{R}=a+b\log_{10}N_\mathrm{ch},
\end{equation}
where $E_\mathrm{R}$ denotes the reconstructed energy.

\section{Energy resolution and systematic uncertainties}
\label{sec:energy-resolution}
In this section, we present details about the accuracy of the reconstructed
energies and the uncertainties that have to be considered for the
reconstruction of the cosmic-ray intensity. The accuracy depends on the
variation of the true shower size caused by the intrinsic shower-to-shower
fluctuations in the atmosphere and also on the accuracy in the reconstruction
of the shower size. 

\subsection{Energy resolution}
\label{energy-resolution}
\begin{figure}
 \includegraphics[width=\columnwidth]{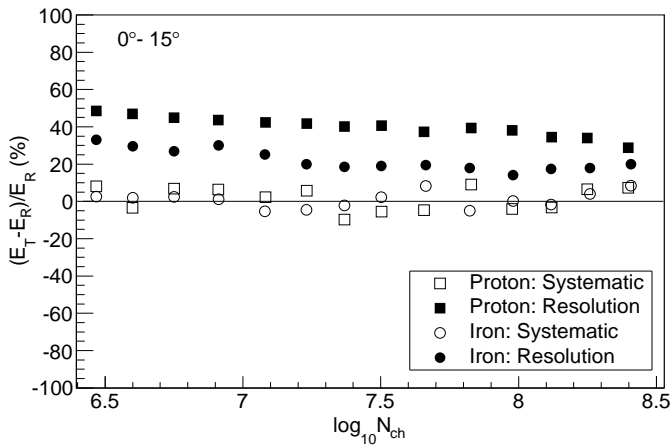}
  \caption{\label {energy-accu} Accuracy in the reconstructed energies
    ($E_\mathrm{R}$) for showers induced by protons (squares) and iron nuclei
    (circles) as a function of the reconstructed size $N_\mathrm{ch}$ for the
    zenith angle bin of $0^\circ-15^\circ$. The filled points represent the
    energy resolution and the the empty points are the systematic
    uncertainties resulting from the energy calibration. See Section \ref{energy-resolution} for details.}
\end{figure}

For each size bin in the $E_\mathrm{T}-N_\mathrm{ch}$ profile plot,
reconstructed energies $(E_\mathrm{R})$ are obtained for every simulated
shower, and a distribution of the differences between the true energies and the
reconstructed energies $(E_\mathrm{T}-E_\mathrm{R})$ is generated. The
distribution obtained is similar to the one shown in Figure \ref{energy-dis},
except for a shift in the peak position to the left by an interval equal to the
value of the reconstructed energy. The peak position and the spread of these
distributions are obtained correspondingly. The peak represents the systematic
uncertainty due to energy calibration, while the spread corresponds to the energy resolution. Their
values expressed as fraction of the reconstructed energies are shown in Figure
\ref{energy-accu} as function of the shower size for the zenith angle range of
$\theta=0^\circ-15^\circ$. The resolution is in the range of $\sim28\%-48\%$
for proton induced showers and $\sim12\%-32\%$ for showers induced by iron
nuclei. The systematics are within 12\% for protons and within 8\% for iron
nuclei. At $\theta=30^\circ-35^\circ$, the uncertainty in energy for protons increases to the range of $\sim37\%-65\%$ in resolution and to $\sim20\%$ in systematics. For iron nuclei, the uncertainty remains almost the same up to $\theta=35^\circ$.

\subsection{Systematic uncertainty in energy}
The systematic uncertainty in energy shown in Figure \ref{energy-accu} is
associated with the energy calibration performed using Equation 
\ref{eq:energy-calibration}. Other main sources of systematic uncertainty in energy include the assumed slope of the primary cosmic-ray spectrum in the CORSIKA simulation, the VEM peak obtained from the detector simulation and the hadronic interaction models. Thus, the calibration parameters, listed in Table
\ref{fit-parameters}, also depend on the choice of simulation parameters.

A part of the systematic uncertainties are obtained by changing the values of
the slope and the VEM peak in the simulations within reasonable limits, and by
comparing the newly reconstructed energies with the energies obtained using the
fixed parameters given in Table \ref{fit-parameters}. For the slope of the
energy spectrum, simulated showers with an original slope
$\gamma_\mathrm{s}=-2$ are weighted to generate distributions for
$\gamma_\mathrm{s}=-2.5$ and $\gamma_\mathrm{s}=-3.5$. Then, following the same
procedure as described in Section \ref{sec:energy-calibration}, energy
calibrations are performed separately for the two different slopes and
calibration parameters are obtained. The differences between the energies
reconstructed with the new parameters and the ones reconstructed using the
parameters given in Table \ref{fit-parameters} gives the systematic uncertainty
due to the spectral slope. The uncertainties are found to be within
$(+6\%,-9\%)$ for protons and within $\pm2\%$ for iron nuclei.

\begin{figure*}
 \includegraphics[width=\columnwidth]{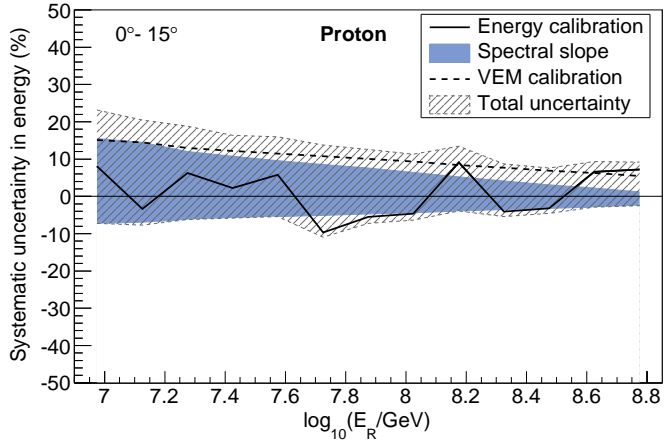}
 \hspace*{\fill}
 \includegraphics[width=\columnwidth]{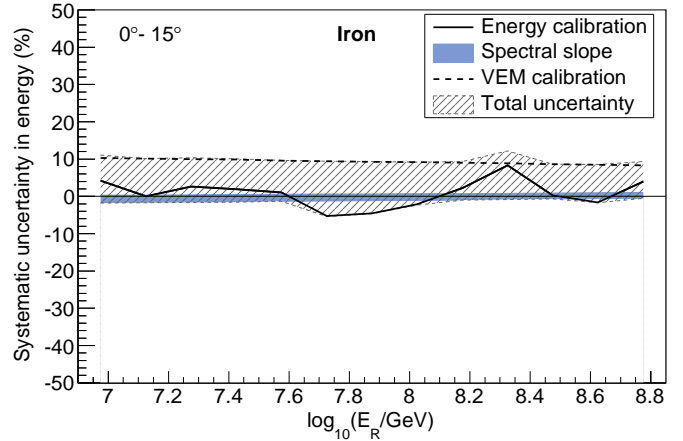}
 \caption{\label {energy-sys} Systematic uncertainties in the reconstructed
    energy $(E_\mathrm{R})$ for showers induced by protons (left panel) and iron
    nuclei (right) for the zenith angle bin of  $0^\circ-15^\circ$ as a
function of $E_\mathrm{R}$.  The systematic uncertainties due to the energy
calibration (thick solid lines) are the same as  shown in Figure
\ref{energy-accu} but plotted as function of $E_\mathrm{R}$. They are
calculated using the parameters set given in Table \ref{fit-parameters} for
$\gamma_\mathrm{s}=-3$.  The blue band represents the uncertainty resulting
from changing the spectral slope from $-2.5$ to $-3.5$.  The dashed line is due
to the uncertainty involved in the VEM calibration and the shaded-striped
region is the total uncertainty.}
\end{figure*}

From the detector simulation, it has been observed that adding noise to the
deposited energy in the detector at the level of $1$~MeV (see Section
\ref{detector-simulation}) leads to around $10\%$ positive shift in the value
of the most probable energy deposition $E_\mathrm{VEM}$ in the detector for
vertical incident muons. The 10\% increase in $E_\mathrm{VEM}$ will lead to a
decrease in the shower size and subsequently to an increase in the
reconstructed energy by $\sim10\%$. The average systematic shift in the
reconstructed energy due to this uncertainty in VEM calibration is obtained to
be $\sim+10\%$ for showers induced by either protons or iron nuclei. The
different systematic uncertainties obtained are shown in Figure
\ref{energy-sys} as a function of the reconstructed energy for showers induced
by protons (left panel) and iron nuclei (right panel). For proton showers, the
total systematic uncertainty, obtained by adding the individual systematic components in quadrature, is found to be within $\sim(+20\%,-10\%)$ and for iron showers, the total systematic is within $(+10\%,-5\%)$. At larger zenith angles, the total systematic for protons increases slightly, reaching $\sim(+22\%,-15\%)$ at $\theta=30^\circ-35^\circ$, while for iron nuclei, the uncertainty remains almost unchanged. 

\subsection{Systematic uncertainty in intensity}
\label{sec:sys-intensity}
Any systematic uncertainty in energy results in a systematic shift in the
reconstructed cosmic-ray flux intensity. To estimate the systematic uncertainty in
intensity due to the energy calibration, the reconstructed energies are
determined using Equation \ref{eq:energy-calibration} for all simulated showers with
$\gamma_\mathrm{s}=-3$ that pass through all selection and quality cuts as
listed in Table~\ref{quality-cut}. The distribution of the reconstructed
energies is compared to the distribution of the true energies, and the
systematic uncertainty in intensity is calculated as
$(I_\mathrm{T}-I_\mathrm{R})/I_\mathrm{R}$ for each energy bin, where
$I_\mathrm{T}$ and $I_\mathrm{R}$ represent the number of showers per bin in
the true and reconstructed energy distributions respectively.

\begin{figure*}
\includegraphics[width=\columnwidth]{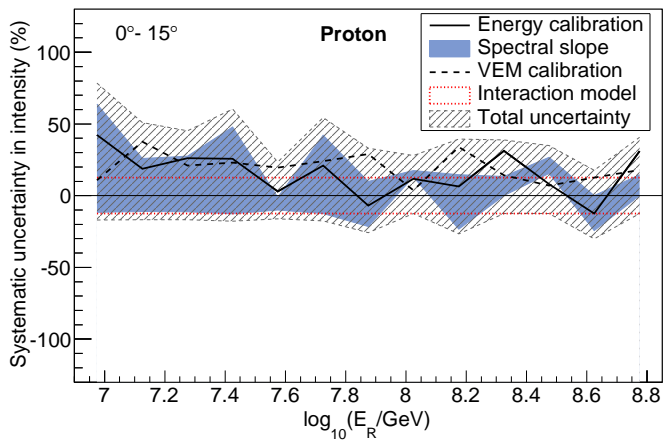}
\hspace*{\fill}
\includegraphics[width=\columnwidth]{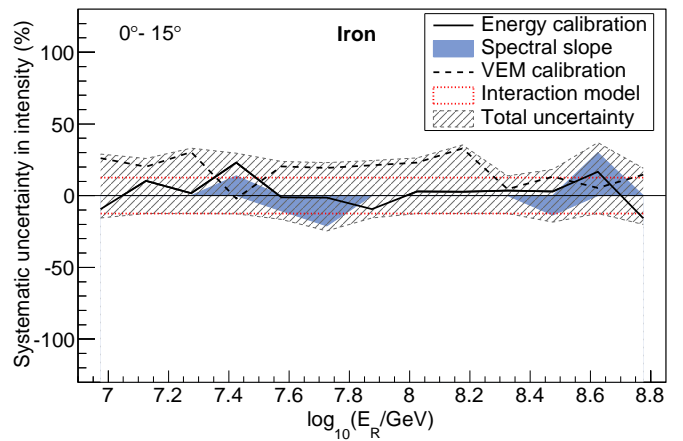}
\caption{\label {intensity-sys} Systematic uncertainties in intensity obtained
from simulations of showers induced by protons (left panel) and iron nuclei
(right panel) for the
zenith angle bin of $0^\circ-15^\circ$. All lines/bands (except the dotted
band) have the same representation as in Figure \ref{energy-sys}. The dotted
band represents the uncertainty expected due to the hadronic interaction model
which is taken as $12.5\%$.}
\end{figure*}

For the systematic effect due to the uncertainties in the spectral slope and
the VEM calibration, the energy calibration determined in their respective
cases are applied to the simulated showers for $\gamma_\mathrm{s}=-3$ and the
distributions of the newly reconstructed energies are compared with the old
distribution obtained using the parameters given in Table \ref{fit-parameters}.

Figure \ref{intensity-sys} shows the different systematic uncertainties in
intensity that have been obtained for protons (left panel) and iron nuclei (right panel). The thick solid line represents the systematic uncertainty due
to the energy calibration, the blue band represents the contribution due to the
spectral slope, the dashed line is the VEM contribution, and the shaded-striped
region represents the total systematic uncertainty. For energies above $\log_{10}(E_\mathrm{R}/\mathrm{GeV})\sim7.2$, the systematic uncertainty due to the energy calibration is found to be within $\sim(+30\%,-10\%)$ for proton showers and within $\sim(+20\%,-10\%)$ for showers induced by iron nuclei.
The systematic uncertainty due to the spectral slope is within
$\sim(+40\%,-15\%)$ for protons, and within $\sim(+12\%,-18\%)$ for iron nuclei except at $\log_{10}(E/\mathrm{GeV})\sim8.6$ where the uncertainty reaches $\sim30\%$. The systematic uncertainty associated with the VEM calibration is found to be within $+30\%$ for both types of nuclei. A contribution of $12.5\%$ due to the uncertainty in the hadronic interaction model \citep{Apel2012, Kang2013} is also included in Figure \ref{intensity-sys}. For protons, the total systematic uncertainty above $\log_{10}(E_\mathrm{R}/\mathrm{GeV})\sim7.2$ is within $\sim(+60\%,-25\%)$, and for iron nuclei, the total uncertainty is within $\sim(+38\%,-20\%)$.

\begin{figure*}
\includegraphics[width=\columnwidth]{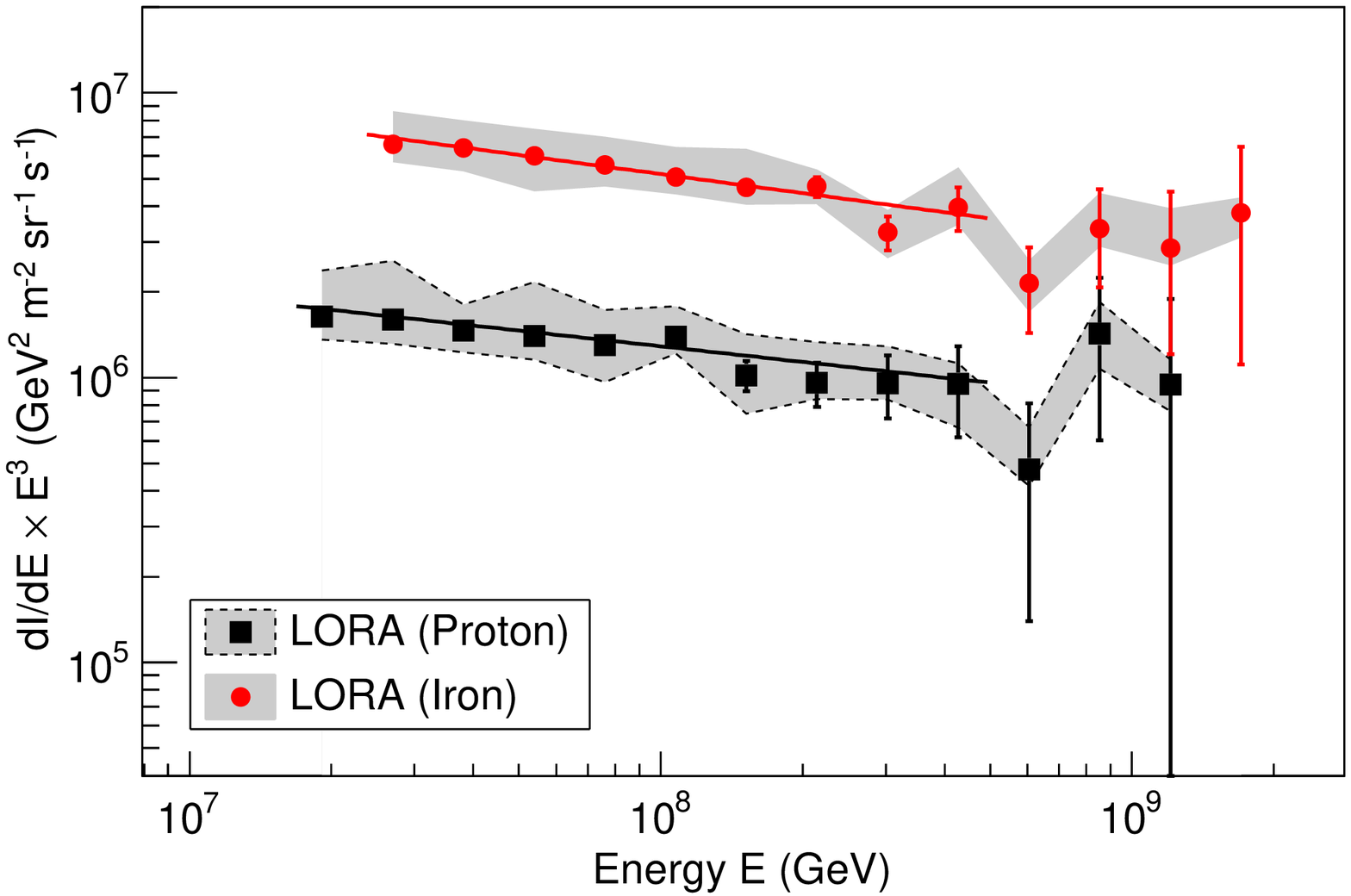}
\hspace*{\fill}
\includegraphics[width=\columnwidth]{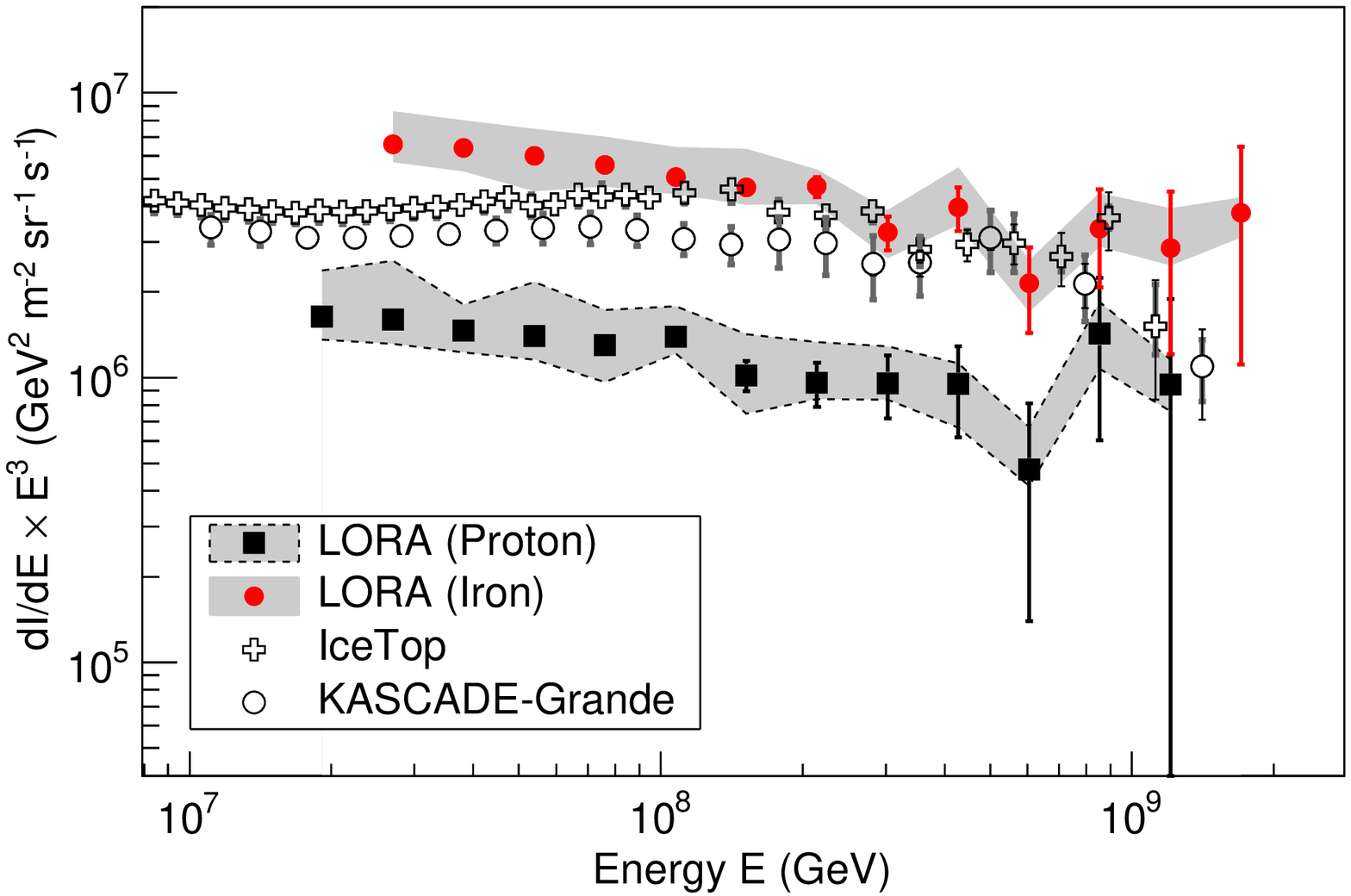}
\caption{\label {cr-spectrum} Left: All-particle cosmic-ray energy spectrum
measured with LORA, assuming that cosmic rays are only protons (squares) and
iron nuclei (filled circles).  The error bars represent statistical
uncertainties and the shaded areas represent systematic uncertainties. The
lines represent single power law fits to the measurements, excluding the
highest three energy bins. 
Right: LORA measurements compared to the all-particle energy spectrum from
IceTop (crosses) and KASCADE-Grande (empty circles) measurements.}
\end{figure*}

\section{Measured cosmic-ray energy spectrum}
\label{sec:cosmic-ray-spectrum}
For all high-quality LORA data, reconstructed energies are determined on
shower-by-shower basis, and a distribution of reconstructed energies is built
taking a logarithmic bin size of 0.15. From the distribution, the differential
cosmic-ray spectrum $(dI/dE)$ is obtained by folding in the total acceptance of
the LORA array $A_\mathrm{acc}$ (Figure \ref{total-effective-area}) and the
total observation time $T_\mathrm{obs}$ as follows,
\begin{equation}
\centering
\left(\frac{dI}{dE}\right)_i=\left(\frac{\Delta n}{\Delta E}\right)_i\times \frac{1}{A_\mathrm{acc}T_\mathrm{obs}}
\end{equation} 
where the subscript $i$ denotes the $i^\mathrm{th}$ energy bin and $\Delta n$
is the number of showers in an energy bin of width $\Delta E$. For constructing
the spectrum, only the energy region that has a total (trigger and
reconstruction) efficiency greater than $98\%$ is used. This corresponds to an
energy of $1.9\times 10^7$~GeV for protons and $2.7\times 10^7$~GeV for iron
nuclei (see Figure \ref{total-efficiency}).

Figure \ref{cr-spectrum} (left panel) shows the reconstructed energy spectrum
multiplied by $E^3$, assuming that cosmic rays are only protons or iron nuclei.
The spectrum is given in the energy range of
$(1.9\times10^{7}-1.2\times10^{9})$~GeV for protons, and in the range of
$(2.7\times10^{7}-1.7\times10^{9})$~GeV for iron nuclei. The measured values
along with the uncertainties are listed in Table \ref{energy-spectra}. The
measured spectra cannot be described by single power laws over the full energy
range because of the structures present in the spectra, particularly the dip at $\sim 6\times 10^8$~GeV. A power law fit to the measured spectra data below $5\times 10^8$~GeV 
gives spectral index values of $\gamma_\mathrm{P}=-3.18\pm0.13$ for protons and
$\gamma_\mathrm{Fe}=-3.22\pm0.08$ for iron nuclei. 

In Figure \ref{cr-spectrum} (right panel), our measured spectra are compared
with the all-particle spectra measured with the IceTop \citep{Aartsen2013} and
KASCADE-Grande \citep{Apel2012} experiments. Both their spectra lie between our reconstructed spectra, which is expected in the case of a mixed cosmic-ray composition. They are close to our proton spectrum at $\sim 2\times 10^7$~GeV, and become closer to our iron spectrum as the energy increases. This might be an indication of a change in the mass composition of cosmic rays in the energy region between $10^7$ and $10^9$ GeV, which is expected as due to a transition from a Galactic to an extragalactic origin of cosmic rays.

\section{Conclusion and outlook}
\label{sec:conclusion}
We have conducted a detailed energy reconstruction study for the extensive air
showers measured with the LORA particle detector array. Important parameters
such as the energy resolution of the array and the systematic uncertainty of
the reconstructed energy have been obtained. The energy resolution is found to
be in the range of $\sim28-48\%$ for showers induced by protons and
$\sim12-32\%$ for iron nuclei.  The total systematic uncertainty of the reconstructed
energy is within $\sim(+20\%,-10\%)$ for protons and within $\sim(+10\%,-5\%)$ for iron nuclei. Applying the
reconstruction method to the measured data, the all-particle cosmic-ray energy
spectrum has been obtained, assuming that cosmic rays are only constituted of
protons or iron nuclei for energies above $\sim10^{16}$ eV with a systematic
uncertainty in intensity of $\sim20-60\%$.  Our future effort will concentrate
on combining the energy measurement of LORA with the composition measurement
from the LOFAR radio antennas to determine an all-particle energy spectrum,
taking into account the actual cosmic-ray composition.

Especially the primary energy determined using the energy calibration given
here is being used in the reconstruction of air shower properties with the
radio data from LOFAR. Calculation of energy calibration parameters for higher
zenith angles above $\sim40^\circ$ is underway. This is particularly important
for the LOFAR radio measurements where a significant fraction of showers
have been observed at larger zenith angles. At present, the small size of the
LORA array effectively limits the effective area of LOFAR. Efforts are ongoing
to expand the size of the array to exploit the full potential of LOFAR.

\begin{table*}
\centering
\caption{Values of the measured cosmic-ray spectrum, assuming that cosmic rays
are only protons or iron nuclei.
The energies are given in GeV and the intensities along with
the statistical and the systematic uncertainties are given in units of
1/(m$^{2}$ sr  s GeV).}
\vspace{\baselineskip}
\label{energy-spectra}
\begin{tabular}{ccc}
\hline
\multicolumn{1}{c}{\rule{0pt}{2.5ex} Energy} & \multicolumn{2}{c}{Intensity $\pm$ stat. $\pm$ sys. uncertainties [1/(m$^{2}$ sr  s GeV)]}\\
\cline{2-3}
\rule{0pt}{2.5ex}
(GeV) & Protons & Iron nuclei\\
\hline
\rule{0pt}{2.5ex}
$1.91\times 10^7$ & $(2.34\pm0.02$ $^{+1.05}_{-0.39})\times 10^{-16}$ & $-$\\
\rule{0pt}{2.5ex}
$2.70\times 10^7$ & $(8.13\pm0.14$ $^{+4.91}_{-1.45})\times 10^{-17}$ & $(3.35\pm0.02$ $^{+0.99}_{-0.42})\times 10^{-16}$\\
\rule{0pt}{2.5ex}
$3.81\times 10^7$ & $(2.63\pm0.06$ $^{+0.61}_{-0.42})\times 10^{-17}$ & $(1.15\pm0.01$ $^{+0.27}_{-0.19})\times 10^{-16}$\\
\rule{0pt}{2.5ex}
$5.39\times 10^7$ & $(8.97\pm0.33$ $^{+4.48}_{-1.58})\times 10^{-18}$ & $(3.84\pm0.06$ $^{+0.88}_{-0.94})\times 10^{-17}$\\
\rule{0pt}{2.5ex}
$7.61\times 10^7$ & $(2.94\pm0.15$ $^{+0.97}_{-0.76})\times 10^{-18}$ & $(1.26\pm0.03$ $^{+0.31}_{-0.19})\times 10^{-17}$\\
\rule{0pt}{2.5ex}
$1.07\times 10^8$ & $(1.11\pm0.08$ $^{+0.31}_{-0.14})\times 10^{-18}$ & $(4.08\pm0.15$ $^{+1.07}_{-0.51})\times 10^{-18}$\\
\rule{0pt}{2.5ex}
$1.52\times 10^8$ & $(2.91\pm0.35$ $^{+1.15}_{-0.77})\times 10^{-19}$ & $(1.32\pm0.07$ $^{+0.47}_{-0.16})\times 10^{-18}$\\
\rule{0pt}{2.5ex}
$2.14\times 10^8$ & $(9.71\pm1.71$ $^{+3.77}_{-1.21})\times 10^{-20}$ & $(4.75\pm0.38$ $^{+0.65}_{-0.59})\times 10^{-19}$\\
\rule{0pt}{2.5ex}
$3.03\times 10^8$ & $(3.43\pm0.86$ $^{+1.20}_{-0.43})\times 10^{-20}$ & $(1.16\pm0.15$ $^{+0.21}_{-0.21})\times 10^{-19}$\\
\rule{0pt}{2.5ex}
$4.28\times 10^8$ & $(1.21\pm0.43$ $^{+0.21}_{-0.36})\times 10^{-20}$ & $(5.06\pm0.88$ $^{+1.87}_{-0.63})\times 10^{-20}$\\
\rule{0pt}{2.5ex}
$6.04\times 10^8$ & $(2.15\pm1.52$ $^{+0.87}_{-0.27})\times 10^{-21}$ & $(9.72\pm3.24$ $^{+1.87}_{-1.95})\times 10^{-21}$\\
\rule{0pt}{2.5ex}
$8.54\times 10^8$ & $(2.28\pm1.32$ $^{+0.66}_{-0.56})\times 10^{-21}$ & $(5.34\pm2.02$ $^{+1.74}_{-0.66})\times 10^{-21}$\\
\rule{0pt}{2.5ex}
$1.20\times 10^9$ & $(5.38\pm5.38$ $^{+1.18}_{-1.05})\times 10^{-22}$ & $(1.62\pm0.94$ $^{+0.59}_{-0.20})\times 10^{-21}$\\
\rule{0pt}{2.5ex}
$1.70\times 10^9$ & $-$ & $(7.66\pm5.42$ $^{+0.97}_{-1.33})\times 10^{-22}$\\
\hline
\end{tabular}
\end{table*}

\section*{Acknowledgment}
We would like to thank the technical support from ASTRON. In particular, we are
grateful to J.~Nijboer, M.~Norden, K.~Stuurwold and H.~Meulman for their
support in the installation and in the maintenance of LORA in the LOFAR core.
We are also grateful to the KASCADE-Grande collaboration for generously lending
us the scintillator units. We acknowledge funding from the Samenwerkingsverband
Noord-Nederland (SNN), the Netherlands Research School for Astronomy (NOVA) and
from the European Research Council (ERC) under the European Unions Seventh
Framework Program (FP/2007-2013) / ERC Grant Agreement no. 227610. LOFAR, the
Low Frequency Array designed and constructed by ASTRON, has facilities in
several countries, that are owned by various parties (each with their own
funding sources), and that are collectively operated by the International LOFAR
Telescope (ILT) foundation under a joint scientific policy.





\end{document}